\documentstyle[11pt]{article}
\begin{document}
\title {\bf A note on light velocity anisotropy}
\author{Bruno Preziosi \and
Dipartimento di Scienze Fisiche - Universit\`a Federico II di Napoli\\
INFM}
\date{}
\maketitle
{\it ABSTRACT}

{\it It is proved that in
experiments on or near the Earth, no anisotropy in the 
one-way velocity of light may be detected. The very accurate experiments
which have been performed to detect such an effect
 are to be considered significant tests
of both special relativity and the equivalence principle}.

{\it Pacs - 03.30 Special relativity 

}

06.30 - Measurement of basic quantities.\\

\vskip 5mm

mailing address:\\
Prof. Bruno Preziosi\\
Dipartimento di Scienze Fisiche, Universit\`a di Napoli "Federico II"\\
Mostra d'Oltremare, Pad.20\\
I-80125 Napoli (Italy)\\
tel.n.: (+39) 81-7253416, fax n.: (+39) 81-2394508\\
e-mail: preziosi@na.infn.it\\

\newpage

\section{Introduction}
\label{s1}

The anisotropy in the 
microwave background \cite{Smo77} has suggested the existence 
of a preferred frame $\Sigma$ which sees an isotropic background and 
of a corresponding anisotropy in the
one-way velocity of light, when measured in our system $S$, which 
moves with
respect to $\Sigma$ at the velocity of about 377 km/s.
Possible consequences have been exploited from the theoretical point of 
view \cite{Rob49} \cite{Man77};
many important and precise experiments have then been 
carried out with the 
purpose of detecting this anisotropy. No variation was 
observed at the level of $3\times10^{-8}$
 \cite{Tur64}, $2 \times 10^{-13}$ \cite{Bri79}, 
$3\times 10^{-9}$ \cite{Rii88}, 
$2 \times 10^{-15}$\cite{Hil90}, $3.5\times 10^{-7}$ \cite{Kri90},
$5\times10^{-9}$ \cite{Wol97}.

Our motion with respect 
to $\Sigma$ is a composition of the motions of the
Earth in the solar system, of this system in our galaxy, of our galaxy
inside a group of galaxies,... .
The problem which arises 
is a very old one: may we perform, on or near the
Earth, experiments to make evident 
our motion with respect to the preferred frame?

Historically, this question has been formulated in two steps,
connected with the relativity principle and the equivalence principle,
respectively.

The first step is due to Galilei, 
who excluded the possibility of performing,
inside a ship cabin, experiments 
having the purpose of measuring the ship
velocity with respect to the mainland. 
To compare the background radiation
case with the Galilei proposal,
if the Sun light entering inside the cabin through a port-hole
is analysed, its
black-body radiation spectrum would appear different from the one
observed on the mainland.

The second step was introduced by Einstein 
through the equivalence principle \cite{Wil93}:
{\it At every space-time point in an arbitrary 
gravitational field it is
possible to choose a "locally inertial coordinate system" 
such that, within a
sufficiently small region of the point in question, 
the laws of nature take the same form 
as in unaccelerated Cartesian coordinate 
systems in absence of
gravitation}.\cite{Wei72}
As a consequence, experiments inside a freely falling space cabin
exhibit its relative motion only in the presence of inhomogeneities
in the gravitational field.

In the following section
 we discuss the quoted experiments according to the two steps
outlined above.

In section 2  we analyse the linear transformations due to
Robertson and to Mansouri and Sexl, which generalize the Lorentz
one and which have stimulated the experiments we are speaking of.
We conclude that, due to the definitional role of  light velocity,
the linear transformations between inertial frames must have
Lorentz form.

In section 3 we analyse the possibility of locally detecting
anisotropies in the light velocity in the case of general relativity.

\section{The preferred  frame}

The linear transformations between inertial frames
 have been analysed by very many authors (a list of them is given 
in ref. \cite{Pre94})
under hypotheses which include the requirement that they form a group,
but do not include, a priori,
the invariance of the light velocity. They conclude that these (Lorentz)
transformations must be characterized by a velocity
$c$, infinite in the Galilei limit, which, in principle, 
may take different absolute values in the different
astronomical directions.

Robertson \cite{Rob49} and 
 Mansouri and Sexl \cite{Man77}
have analysed the linear transformation between
the preferred reference frame $\Sigma$ 
and another inertial frame $S$ which is moving with respect to it.

Robertson derives the following general linear transformation between 
the preferred system $\Sigma(x',y',z',t')$ 
and the frame $S(x,y,z,t)$, which
moves along the $x$-direction which connects the respective origins $\Omega$
and $O$:
\begin{equation}
x'=a_{1}x+va_{0}t,~~~~y'=a_{2}y,~~~~z'=a_{2}z,~~~~t'=\frac{va_{1}}{c^{2}}x
+a_{0}t,
\end{equation}
where $a_{0}$, $a_{1}$ and $a_{2}$ may depend on $v$.
This transformation, which is expressed in terms of the parameter $v$
and which reduces to the identity when $v=0$,
 is derived under the hypotheses that:

 $i)$ space is euclidean
for both $\Sigma$ and $S$;

$ii)$ in
$\Sigma$ all clocks are synchronized and light moves 
with a speed $c$ which
is independent of direction and position;

 $iii)$ the one-way speed of
light in $S$ in planes perpendicular to the motion of $S$
 is orientation-independent.

Notice that $O$ moves with respect to $\Sigma$ with
velocity $v$, while  $\Omega$ moves with respect to $S$ with
velocity $\tilde{v}\equiv v a_{0}/a_{1}$ ; analogously, light,
which is seen 
by $\Sigma$ to move according to the law $x'=ct'$, is seen by
$S$ to move with velocity
$\tilde{c}\equiv c a_{0}/a_{1}$, independently of 
the direction of $c$. If $c$ is the maximum speed in $\Sigma$, $\tilde{c}$ 
is the maximum speed in $S$; moreover $\tilde{v}/\tilde{c}=v/c$.
In terms of these
true velocities, equations (1) take the form
\begin{equation}
t'=a_{0}\left(t+\frac{\tilde{v}}{\tilde{c}^{2}}x\right),~
x'=a_{1}\left(x+\tilde{v}t\right),~
y'=a_{2}y,~
z'=a_{2}z.\nonumber\\
\end{equation}
The transformation is then the product of a Lorentz transformation and
a scale transformation; the latter may be re-absorbed by a redefinition of the
units.

If the length standard is estabilished, in any frame,
by giving an assigned value to the speed of
light, then the light velocities in $\Sigma$ and $S$ are equal, $x$
is scaled by $a_{0}/a_{1}$ and the transformation between the $(x,t)$
variables  takes a familiar form.

 The fact that this transformation implies different
light speeds in different directions in the $(x,y)$ plane is, a priori,
admissible.

This case is typical of a tetragonal crystal; the light speeds may be different
in the $x$ and in the $y$ directions, when measured with external
standards;  a suitable internal scaling of the $y$ variable would of
course give the same value to the internal velocities, but the time
required for the light to travel the crystal in the $y$-direction would
be different from the one seen by an external observer.
In this case the attribution of different light speeds for different
directions is physically justified. But if we have no such a justification,
the thing to do is to apply the Poincar\'e simplicity criterion and to
attribute the same value to the light speed in different directions.

Mansouri and Sexl \cite{Man77} 
 analyse the linear 
transformation from a preferred frame $\Sigma (X,T)$ to
 another frame 
$S (x,t)$, which moves with respect to it at the velocity $v$,
 under the hypothesis that 
the synchronization is realized by clock transport.
Their analysis is devoted both to one-dimensional transformations and
to three-dimensional ones; we do not discuss here the last case,  
but the conclusions will apply as well.

 To first order in $v$, the Mansouri and Sexl
one-dimensional transformation takes the form:
\begin{equation}
\left(\begin{array}{c}x\\t\end{array}\right)=
\frac{1}{\sqrt{1-(1-\mu)\frac{v^{2}}{c^{2}}}}\left(\begin{array}{cc}
1 & -v\\ -\frac{(1-\mu) v}{c^{2}}&1
\end{array}\right)
\left(\begin{array}{c}X\\T\end{array}\right),
\end{equation}
where $c$ is the isotropic light speed seen by $\Sigma$
and their quantity $2\alpha$ 
is substituted here by $-(1-\mu)/c^{2}$ to make explicit the Lorentz and
Galilei limits 
($\mu=0$ and $\mu=1$, respectively). 
If $\mu\ne 0$, then a particle, which moves with velocity
 $u$ in $\Sigma$, appears to $S$ to move according the law:
\begin{equation}
x=\frac{u-v}{1-(1-\mu)\frac{uv}{c^{2}}}t.
\end{equation}
The maximum speed of a frame with respect to $\Sigma$
is $c/\sqrt{1-\mu}$,
 independently of the orientation; if something  is seen by
$\Sigma$ to move at this speed, it is seen to move
at the same invariant speed by all frames, independently of the orientation.
On the other hand, light, which moves with respect to
$\Sigma$ according to $X=\pm cT$, moves, in our frame $S$,
 according to $x=t(c\mp v)/(1\mp (1-\mu)v/c^{2})$,
the sign depending on the motion orientation.
Light speed is no more the maximum one, and, what is more relevant,
the one-way light velocities in $S$ are different.

The undetectability of a possible dependence of
the one-way velocity along a line on its orientation has been extensively
discussed in literature \cite{Rei28} \cite{Gru73}. As a consequence, $\mu=0$. 

The last indisputable conclusion finds a confirmation in the following
experimental fact: electrons in the large accelerator machines have now 
energies of $\sim 100$ Gev; at this energy,
$\frac{v^{2}}{c^{2}}\sim 1-2\cdot 10^{-11}$, but the
 electrons have not reached the light
speed; we must have then 
$\mu<2\cdot 10^{-11}$.

The situation does not change if we go beyond the lowest order and suppose
that $\mu$ is a function (even) of $v$.

\section {Local inertial systems}

The above considerations refer to situations in which we are performing
our experiments in frames which are seen by $\Sigma$ to move inertially.
Our conclusion is that, if the transformation between the preferred
frame and our one is taken to be linear, then it must have Lorentz form.

On the other hand, the anisotropy of the primordial radiation strongly 
supports the existence of the preferred frame with respect to which
we are moving. It must  then be analysed
how our state of non-inertial motion affects the experiments we are 
discussing. 

The starting point is the fact that 
the background radiation intensity appears to  be anisotropic to
an observer $O$,
at the origin of a reference frame $S$ in our region $R$ of the universe,
while it is isotropic from the point of view of an observer $\Omega$, at 
the origin of a preferred reference frame $\Sigma$.

In the last case, the absolute 
system $\Sigma$ and the  relative
frame $S$ of our region of the universe detect differences in the
radiation background, but no differences in any {\it local}
experiment.
 The region $R$ behaves like the world 
inside an Einstein elevator; the Einstein equivalence 
principle states that,
if $\Omega$ and $O$ perform, in their 
respective regions, identical experiments  
which are not influenced by the presence of local masses (Earth, 
Sun, ...), they obtain identical results.
 An immediate consequence is
that the inertia principle is valid for all  local inertial systems.

This concept is very clearly stated by Hans Reichenbach \cite{Rei28}:

{\it According to Einstein, however, only these
local systems are the actual inertial systems. In them the field, which 
generally consists of a gravitational and an inertial component, is
transformed in such manner that the gravitational component disappears
and only the inertial component 
remains.}

Analogously, the local inertial systems associated to an Einstein
elevator are connected by linear transformations characterized by
an invariant velocity $c$.
So, our region $R$ and another region $R'$ in the universe have  
separate families of local inertial frames, characterized by identical light 
speeds, although these ones
may appear different when measured by an asymptotic observer
who sees how space curvature modifies in going from $R$ to $R'$.
 A well known example is given by the time delay measured in the Shapiro
\cite {Sha71} and Reasenberg \cite{Rea79} experiments: an asymptotic
observer detects a delay in the light trip, but any observer, who is
in the region this ray is passing through, says that it is moving at the
speed of $c$.

Coming back to the experiments performed in presence of Earth and Sun, 
we do not exclude that local observers may see general
relativistic effects induced by their masses \cite{Cha83}.
The light behaviour is, however,
locally influenced by the gravity only for a
bending which is  very small and difficult to detect \cite{Pre98}; 
this is not true for the motions
of the satellytes involved in some of the quoted experiments.
All gravity effects due to the nearest relevant masses
 have been consistently taken into account in the previous experiments,
which must be highly considered for their precision.

The conclusions of these experiments is
 that, apart from some very small local effects,
the  Lorentz transformation applies in our region $R$, and that 
that our region belongs to a family of local inertial frames.

The force which induces the acceleration seen by some asymptotic observer
is completely cancelled by the equivalence principle. 

\section{Conclusions}
If there is no way to perform
independent measures of lengths and light velocities; in other words,
if the light velocity is used both for synchronizing clocks and for
fixing the unit lengths, 
there is no way of locally detecting any dependence on the orientation of
the length of) a rod. The only thing to do is to use the Poincar\'e
simplicity criterion and consider equal the lengths of the rods and the
one-way speeds of light in the different directions.

In conclusion, isotropy in the one-way velocity of light is a matter of
definition. 

However,
the experiments quoted at the beginning, in particular those
performed by J. Hall and coworkers at very 
sophisticated levels, cannot be considered 
simply significant improvements of
classical special relativity tests. 

As discussed in the introduction,
this would surely be the case
if the quoted experiments had been performed in a region where
gravity effects are compensated. But the presence of an anisotropic
background radiation, when interpreted as testimonial of an
analogous anisotropic mass distribution, and the fact that these
experiments find their explanation in the frame of the special
relativity, strongly support the equivalence principle.

In conclusion we strongly suggest that the
accurate conclusions of such experiments
should be considered significant tests of both special relativity 
and the equivalence principle. 

\section{Acknowledgments}
Thanks are due to professors John L. Hall and Giuseppe Marmo for useful 
discussions; we are indebted too with G.Marmo and G. Esposito for a
critical reading of the manuscript. Thanks are also due to an anonymous
referee  who has helped in improving the presentation of  the paper.

\end{document}